%% file: bare_jrnl_new_sample4.tex
\documentclass[lettersize,journal]{IEEEtran}
\usepackage{amsmath,amsfonts}
\usepackage{algorithmic}
\usepackage{algorithm}
\usepackage{array}
\usepackage[caption=false,font=normalsize,labelfont=sf,textfont=sf]{subfig}
\usepackage{textcomp}
\usepackage{stfloats}
\usepackage{url}
\usepackage{verbatim}
\usepackage{graphicx}
\usepackage{siunitx}
\usepackage{xcolor}
\usepackage{cite}
\usepackage[nolist,nohyperlinks]{acronym}
\begin{document}
\input{acro}
%\title{A Differential Bitcell Based on the Ferroelectric Field-Effect Transistor}
\title{A Novel FeFET Differential Bit-Cell With Hybrid Volatile and Non-Volatile Memory Modes}

\author{Jianze Wang, Wei Zhang and Xuanyao Fong \\
Department of ECE, National University of Singapore, 117583 Singapore \\
Email: jianze.wang@u.nus.edu, zhwei98@nus.edu.sg, kelvin.xy.fong@nus.edu.sg

}

\maketitle

\begin{abstract}
% A differential memory bit-cell consisting of a pair of cross-coupled ferroelectric field-effect transistors and a pair of access transistors is proposed.
% %The proposed differential bit-cell consists of two cross-coupled FeFETs and two access transistors.
% The bit-cell can be configured as both volatile mode and non-volatile mode by adjusting the write condition.
% As non-volatile SRAM, the store power of the proposed bit-cell is 0.13~\si{\mu W} with 2~ns store time, which is significantly reduced compared with previous designs.
% Furthermore, explicit Backup and Restore operations are not required.
Non-volatile SRAM (nvSRAM) designs have been investigated to address the high leakage power of CMOS-based SRAM and the large write latency of emerging non-volatile memory (eNVM) technologies. However, prior nvSRAM designs that combine SRAM with eNVM devices typically require backup and restore (B\&R) operations and incur significant cell-area overhead. Here, we propose a differential memory bit-cell consisting of a pair of cross-coupled ferroelectric field-effect transistors (FeFETs) and a pair of access transistors, resulting in a four-transistor (4T) structure, which is smaller than conventional 6T SRAM and many prior nvSRAM designs. The proposed bit-cell can be configured to operate in either volatile or non-volatile mode by adjusting the write conditions. In the non-volatile mode, the proposed nvSRAM achieves a store power of 0.13~\si{\mu}W with a 2~ns store time, and no explicit B\&R operation is required. The proposed bit-cell can also be viewed as a cross-coupled gain cell, enabling further applications.
\end{abstract}

\begin{IEEEkeywords}
FeFET bit-cell, non-volatile SRAM, non-volatile ferroelectric memory
\end{IEEEkeywords}

\input{intro}

\input{background}

\input{memory}

\input{performance}

\input{conclusion}

\bibliographystyle{IEEEtran}
\bibliography{reference}

\vfill

\end{document}

%% file: acro.tex
\newacro{FE}{ferroelectric}
\newacro{IoT}{internet of things}
\newacro{FeFET}{ferroelectric field-effect transistor}
\newacro{eNVM}{emerging non-volatile memory}
\newacro{MW}{memory window}
\newacro{SOI}{silicon-on-insulator}
\newacro{IL}{insulator layer}
\newacro{WL}{wordline}
\newacro{BL}{bitline}
\newacro{SL}{source line}
\newacro{RDF}{random dopant fluctuation}
\newacro{MC}{Monte Carlo}
\newacro{QoI}{quantity of interest}
\newacro{NLS}{nucleation-limited switching}
\newacro{PCC}{Pearson correlation coefficient}
\newacro{nvSRAM}{non-volatile SRAM}

%% file: intro.tex
\section{Introduction}
\label{sec:intro}

%\IEEEPARstart{R}{ecent} \ac{IoT} and AI applications have driven the demand for low-power and high-performance \ac{eNVM} technologies~\cite{nvimcfefet}.
%The power supply for \ac{IoT} applications and devices is usually limited and have normally-off and instantly-on operation patterns.
%Although SRAM is widely used in the embedded memory of processors and in the \ac{IoT} applications, their leakage current issue becomes more severe as the technology node scales down and causes undesired power consumption.

\IEEEPARstart{S}{everal} \ac{eNVM} techonologies, such as RRAM, PCM, and STT-MRAM have been proposed to replace the SRAM as the embedded memory for the applications mentioned above due to their normally-off feature~\cite{reramtcam}.
Although the leakage power can be reduced compared with CMOS-based SRAM, \acp{eNVM} still face challenges such as high write energy consumption, large write latency, and limited endurance.
To overcome these challenges, \ac{nvSRAM}, which is a hybrid memory bit-cell design that combines conventional CMOS SRAM and \acp{eNVM} to gain advantages from both SRAM and \acp{eNVM}, has been studied.
A common design approach for \ac{nvSRAM} is to connect the \ac{eNVM}s to the storage nodes of the latch in the SRAM bit-cell~\cite{nvsram8t}.
However, these designs require explicit Backup and Restore (B\&R) operations to store and restore the state of the latch into and from the \ac{eNVM} devices when the system is entering and exiting the sleep state, respectively.
Thus, an \ac{nvSRAM} that has significantly reduced energy consumption and delay for B\&R operations ($\textit{E}_\text{B\&R}$ and $\textit{t}_\text{B\&R}$, respectively) is desirable for \ac{IoT} applications.
%One critical Figure-of-Merit of \ac{nvSRAM} is the energy consumption of Backup and Restore (B\&R) operation, i.e. $E_\text{B\&R}$, which is the energy consumed by the nvSRAM bit-cell to backup the state of the SRAM to the \ac{eNVM}s and restore the state from \ac{eNVM}s to the SRAM nodes.
%This design takes advantage of the less B\&R operation than normal write operation in recent \ac{IoT} and AI applications, therefore achieving comparable write speed with CMOS SRAM and lower energy consumption as \ac{eNVM}s.

The \ac{FeFET} is a promising \ac{eNVM} that has gained much research attention recently due to its low power consumption, high speed, and high endurance~\cite{nvimcfefet,transpose2024}.
Although \ac{FeFET}-based \ac{nvSRAM} has been proposed to improve $E_\text{B\&R}$ and further reduce the energy consumption, the existing design has large area overhead~\cite{nvsramfefet}.
A 4T-reconfigurable \ac{FeFET}-based differential nvSRAM bit-cell design was proposed, which shows comparable performance to other \ac{nvSRAM}~\cite{nvsram4t}.
However, the reconfigurable \ac{FeFET} has a complicated fabrication process because it requires two separate ferroelectric gate stacks over the channel.
In this work, we propose a 4T differential bit-cell based on conventional \ac{FeFET} without any additional requirements of the fabrication process and achieves comparable performance with CMOS SRAM and other \ac{nvSRAM}.
Moreover, explicit B\&R operations are not required.

The rest of this paper is structured as follows.
First, Section~\ref{sec:background} presents the basics of the \ac{FeFET}.
Then, we present the 4T differential bit-cell design and operations in Section~\ref{sec:memory}.
Section~\ref{sec:performance} discusses the evaluation and analysis of our proposed bit-cell before conclusions are drawn.

\begin{figure}[!b]
    \centerline{\includegraphics[width=\linewidth]{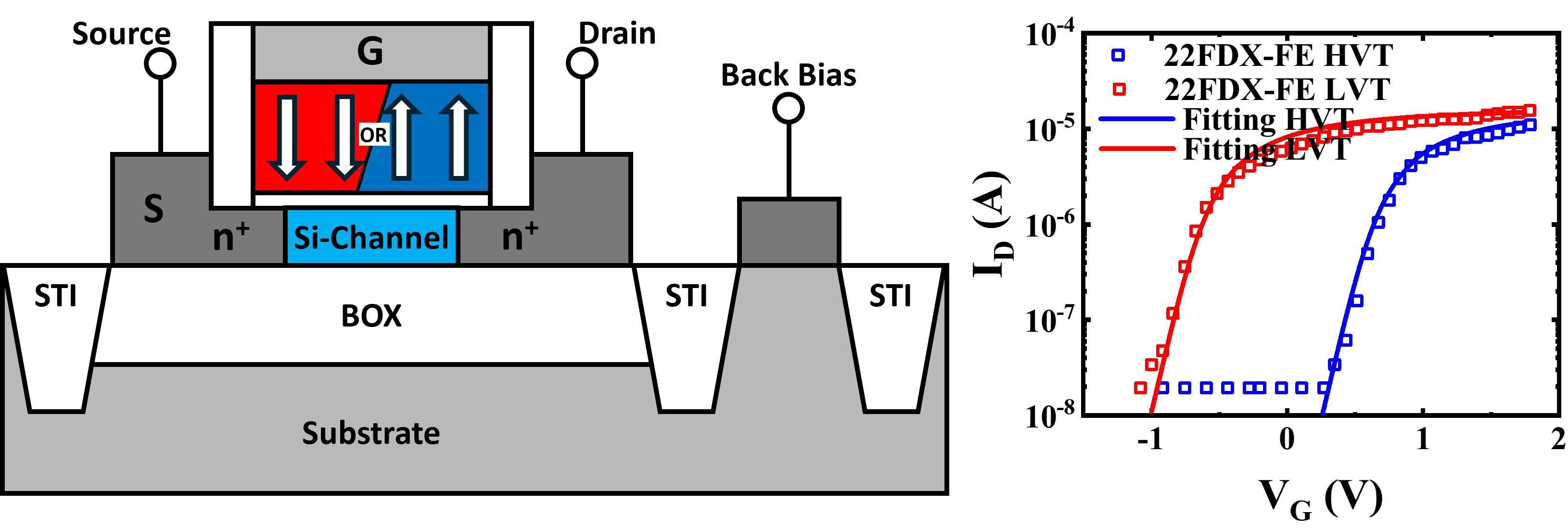}}
    \caption{An FDSOI FeFET for non-volatile memory and the same structure is used for simulation in the following section, the $I$-$V$ of FeFET is calibrated with experimental data from \cite{22fdxexp}.}
    \label{fig:fefet-background}
\end{figure}

%% file: background.tex
\section{Preliminaries}
\label{sec:background}
The cross-section of a \ac{FeFET} bit-cell with a planar \acf{SOI} structure is shown in Fig.~\ref{fig:fefet-background}. 
The \ac{FE} HZO layer is sandwiched between the gate metal and the insulator layer.
% The gate metal is used as the \acf{WL} whereas the drain and source terminals of the \ac{FeFET} are connected to the \acf{BL} and the \acf{SL}, respectively.
The back bias is connected to the substrate of the \ac{FeFET}, which can be used to adjust the threshold voltage, $V_\text{th}$, of the \ac{FeFET}.

In the \ac{FE} layer, its polarization is represented by an arrow that indicates the general orientation of the \ac{FE} dipoles.
As shown in Fig.~\ref{fig:fefet-background}, there are positive charges near the \ac{FE}-\ac{IL} interface when the polarization of the \ac{FE} layer is pointing down (negative polarization, as the red part in the figure), which lowers the $\textit{V}_\text{th}$ to $\textit{V}_\text{th,L}$.
On the other hand, the $\textit{V}_\text{th}$ of the \ac{FeFET} is increased to $\textit{V}_\text{th,H}$ when the polarization of the \ac{FE} layer points up (positive polarization, as the blue part in Fig.~\ref{fig:fefet-background}).
In the conventional \ac{FeFET} bit-cell~\cite{22fdxexp}, its \acf{MW} is the difference between $\textit{V}_\text{th,H}$ and $\textit{V}_\text{th,L}$.
Logic ``1'' and ``0'' are represented by setting the state of the \ac{FeFET} to $\textit{V}_\text{th,L}$ and $\textit{V}_\text{th,H}$, respectively.

The \textit{I}-\textit{V} transfer curves for a \ac{FeFET} are shown in Fig.~\ref{fig:fefet-background}.
In this work, the \ac{FE} layer is modeled as a \ac{FE} capacitor (FeCap) using the \ac{NLS} model~\cite{nlsfefet,EDTM}.
The model for the \ac{FeFET} is obtained by combining the FeCap model and a BSIM-IMG model \cite{bsimimg}.
Fig.~\ref{fig:fefet-background} shows that our \ac{FeFET} model is well-calibrated to the data reported in~\cite{22fdxexp}.

%% file: memory.tex
\section{Proposed 4T Differential Bit-cell}
\label{sec:memory}

\subsection{Bit-Cell Design}
\begin{figure}[!b]
    \centerline{\includegraphics[width=\linewidth]{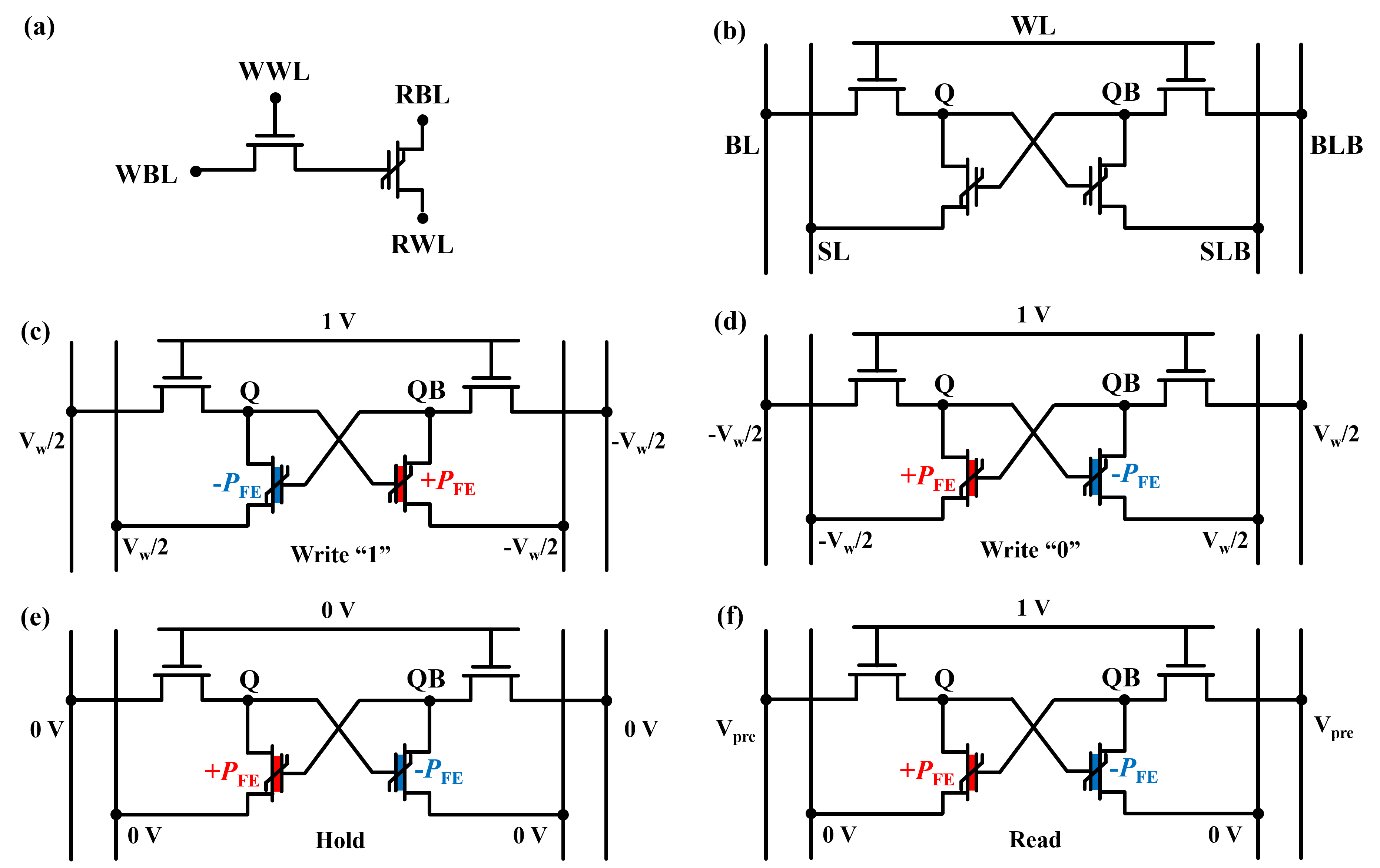}}
    \caption{(a) Conventional 1T-1FeFET gain cell from \cite{1t1fe}. (b) Proposed 4T differential bit-cell design, the bit-cell consists two access transistors, N1 and N2, and two cross-coupled \acp{FeFET}, NF1 and NF2, forming a cross-coupled gain cell. (c) The voltages applied to the bit-cell to write logic ``1''. (d) The voltages applied to the bit-cell to write logic ``0''. (e) Hold operation of the bit-cell. (f) Read operation of the bit-cell.}
    \label{fig:bit-cell}
\end{figure}

The 1T-1FeFET (2T) gain cell from \cite{1t1fe} is shown in Fig.~\ref{fig:bit-cell}(a). Our proposed 4T differential memory bit-cell consists of a pair of the 2T gain cells connected in a cross-coupled topology as shown in Fig.~\ref{fig:bit-cell}(b). As Fig.~\ref{fig:bit-cell}(b) shows, our cross-coupled FeFET gain cell consists of two access transistors (N1 and N2) and two cross-coupled FeFETs (NF1 and NF2).
The gates of N1 and N2 are connected to the \ac{WL} for bit-cell selection and passing the charge on $Q$ and $QB$ to the differential bitlines, BL and BLB.
The cross-coupled \acp{FeFET} are used to store the state of the bit-cell.
The sources of the \acp{FeFET} are connected to the differential source lines, SL and SLB, to stabilize write operation of the \acp{FeFET} and reduce the write energy.
% \textcolor{blue}{Compared with conventional 2T gain cell (1T-1FeFET gain cell from \cite{1t1fe} shown in Fig.~\ref{fig:bit-cell}(a)), the proposed bit-cell constructs a cross-coupled structure of two 1T-1FeFET gain cell. Therefore, the proposed bit-cell can be regarded as a cross-coupled gain cell.}

\begin{figure}[!b]
    \centerline{\includegraphics[width=\linewidth]{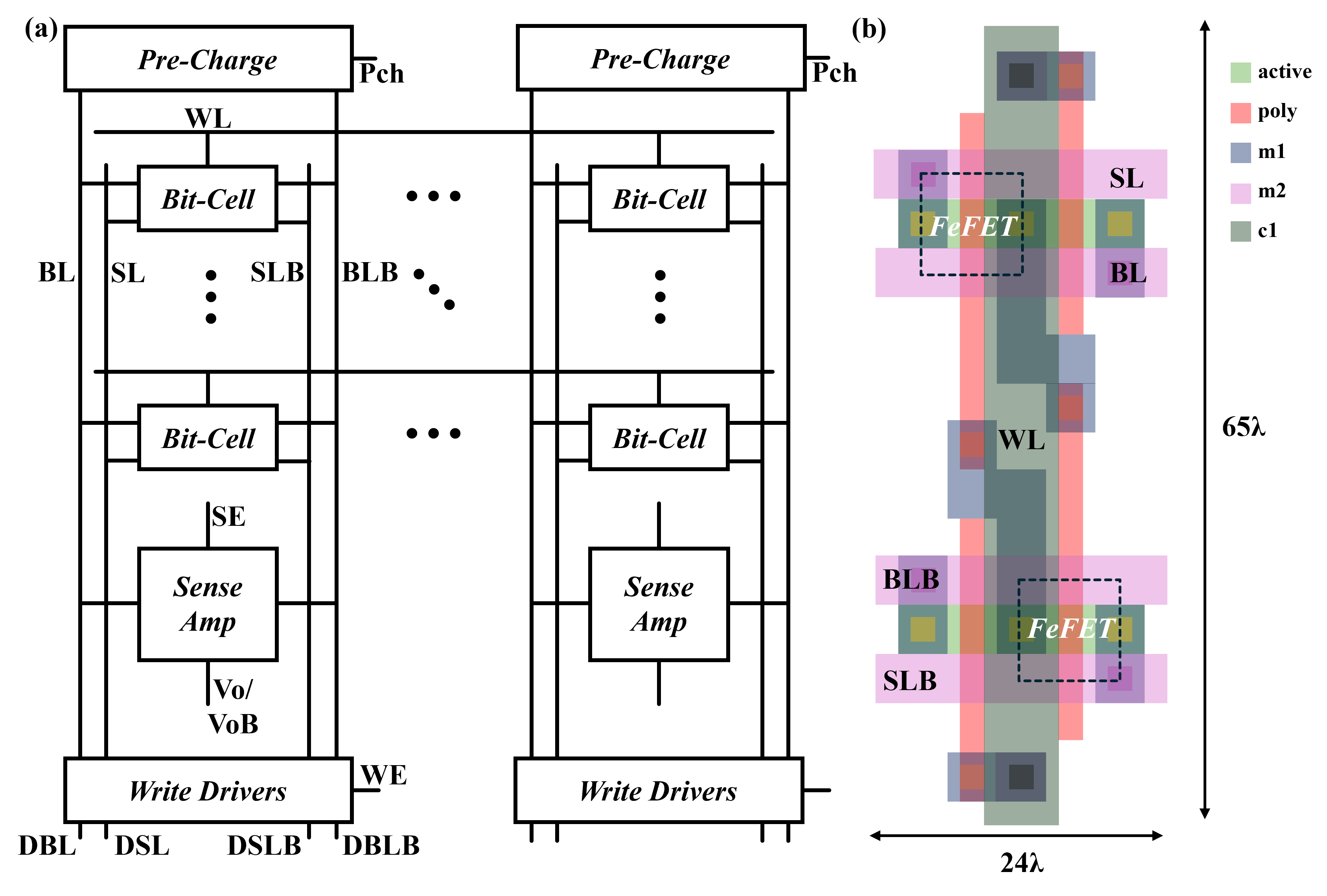}}
    \caption{(a) Proposed non-volatile 4T differential memory array with peripheral circuits. (b) Layout of the proposed 4T differential memory bit-cell.}
    \label{fig:array-and-layout}
\end{figure}

The write operations for our proposed bit-cell are shown in Fig.~\ref{fig:bit-cell}(c) and (d).
To write logic ``1'' into the bit-cell, the polarization of \ac{FE} in NF1 and NF2 need to be $-P_\text{FE}$ and $+P_\text{FE}$, respectively.
The voltages of BL and SL, $\textit{V}_\text{BL}=\textit{V}_\text{SL}=+\textit{V}_\text{w}/2$ and the voltages of both BLB and SLB, $\textit{V}_\text{BLB}=\textit{V}_\text{SLB}=-V_\text{w}/2$ to create a $-V_\text{w}$ voltage drop at $V_\text{gs}$ on NF1 and a $+V_\text{w}$ voltage drop at $V_\text{gs}$ on NF2, as shown in Fig.~\ref{fig:bit-cell}(c).
To write logic ``0'' to the bit-cell, the polarization of \ac{FE} in the cross-coupled \ac{FeFET}s needs to be reversed from Fig.~\ref{fig:bit-cell}(c)(\textit{i.e.} $+P_\text{FE}$ in NF1 and $-P_\text{FE}$ in NF2), and $\textit{V}_\text{BL}$, $\textit{V}_\text{BLB}$, $\textit{V}_\text{SL}$, and $\textit{V}_\text{SLB}$ are as shown in Fig.~\ref{fig:bit-cell}(d).
Note that $\textit{V}_\text{BL}=\textit{V}_\text{SL}$, and $\textit{V}_\text{BLB}=\textit{V}_\text{SLB}$ during write operation.
Thus, current flowing through the bit-cell during write operation is minimized, which reduces the write energy.
In the Hold operation shown in Fig.~\ref{fig:bit-cell}(e), no voltages are applied to BL, BLB, SL, and SLB, which is also the condition during normally-off operation.
The memory state is stored as the non-volatile polarization of the \ac{FE} layers inside the cross-coupled \acp{FeFET}.
Therefore, when used as \ac{nvSRAM}, no explicit B\&R operation is required to backup the state or to restore the state of the bit-cell in contrast to previous \ac{nvSRAM}.
The read operation, illustrated in Fig.~\ref{fig:bit-cell}(f), begins with both BL and BLB precharged to $V_\text{pre}$. When WL is activated, BL and BLB are discharged through NF1 and NF2, respectively. Due to the $V_\text{th}$ difference between NF1 and NF2, the discharge rates differ, enabling the sense amplifier to differentially detect the stored data.

The proposed bit-cell employs HZO for its CMOS compatibility, but its circuit functionality is independent of the specific ferroelectric material as long as the access transistors are carefully designed to be compatible with the required write voltage level. Using other ferroelectrics (e.g., PZT) would mainly alter the required write voltage while preserving the same operational behavior.
It should be noted that the proposed bit-cell requires negative write voltages to reach the desired states, which may increase the complexity of the peripheral circuitry.

\subsection{Memory Array and Layout}

% Fig.~\ref{fig:array-and-layout} (a) shows all components for the 4T differential memory design in our simulations. 
% The blue shaded block is the pre-charge circuit for read operaiton.
% PCH is the enable signal for the pre-chagre circuit for pre-charging the \ac{BL} and BLB, and equalize \ac{BL} and BLB.
% The pink block in the middle is the bit-cell of the 4T differential memory.
% Below the bit-cell block is the latch-based sense amplifier which is commonly used in SRAM, shaded in the light green block.
% SE is the sense enable signal to enable the sense amplifier, Vo and VoB are the output signals from the sense amplifier.
% At the bottom is the write drivers shaded as light grey block.
% WE is the write enable signal for write drivers.
% DBL, DSL, DBLB, and DSLB are the input signals to the write drivers.
Fig.~\ref{fig:array-and-layout}(a) shows the proposed 4T differential memory array with the peripheral circuits.
The bit-cells in the same row are connected through \ac{WL} and the bit-cells in the same column share the same BL, SL, BLB and SLB.
At the top of the array is the pre-charge circuit for read operation.
Pch is the enable signal of the pre-charge block.
A latch-bashed sense amplifier~\cite{senseamp} like that commonly used in SRAM is placed at the bottom of each column for differential sensing of the bit-cell.
% SE for sense enable, Vo and VoB are the outputs from the sense amplifier.
The SE signal is used to enable the sense amplifier (SA) whereas Vo and VoB are the outputs of the SA.
% The data is sent through the write drivers to separate the memory array and the external signals.
WE (write enable) the enable signal of the write drivers and DBL, DSL, DBLB, and DSLB are the input signals to the write drivers.
Except for the use of negative write voltages and differential SL/SLB control, the proposed bit-cell shares the same peripheral scheme as conventional SRAM. The peripheral design details are considered beyond the scope of this paper.

The layout of the proposed 4T differential \ac{FeFET} bit-cell is implemented using GLOBALFOUNDRIES 22~nm PDK with $\lambda$-rule, shown in Fig.~\ref{fig:array-and-layout} (b).
The width and height of the proposed bit-cell are 24$\lambda$ and 65$\lambda$, respectively, resulting in a total bit-cell area of 1560$\lambda^{2}$.
%The bit-cell area can be further optimized to improve the area efficiency.
For the simulations in the later sections, parasitic extraction was performed on the \ac{FeFET} bit-cell to model circuit parasitics in the bit-cell.
%The parasitic parameters are extracted from this layout and are included in the bit-cell simulations.

% \subsection{Simulation Waveforms}

% Fig.~\ref{fig:signals} shows the signals waveform applied to the 4T differential memory in our simulation.

% \begin{figure}[!t]
%     \centerline{\includegraphics[width=0.9\linewidth]{images/waveform.png}}
%     \caption{Waveform of signals applied to the memory circuit in Fig.~\ref{fig:memory-circuit}. The bit-cell will first be written to ``1'' followed by read operation and then written to ``0'' followed by another read operation.}
%     \label{fig:signals}
% \end{figure}

% The bit-cell go through two write-read cycles.
% As the figure shows, in the first cycle, the bit-cell is written to ``1'' by applying $V_\text{w}/2$ pulses on DBL and DSL and $-V_\text{w}/2$ pulses on DBLB and DSLB.
% This write operation is followed by a read operation: PCH is lowering to GND to pre-charge and equalize the \ac{BL} and BLB, then SE is on to enable the sense amplifier, Vo and VoB will output the corresponding voltages that read from the 4T differential memory bit-cell.
% After the first write-read cycle complete, the bit-cell go through the second cycle by writing the cell to ``0'', i.e. $-V_\text{w}/2$ pulses on DBL and DSL and $V_\text{w}/2$ pulses on DBLB and DSLB.
% The read operation is the same as the first cycle.

%% file: performance.tex
\section{Evaulation and Analysis}
\label{sec:performance}

\subsection{Transient Demonstration}

\begin{figure}[!t]
    \centerline{\includegraphics[width=0.9\linewidth]{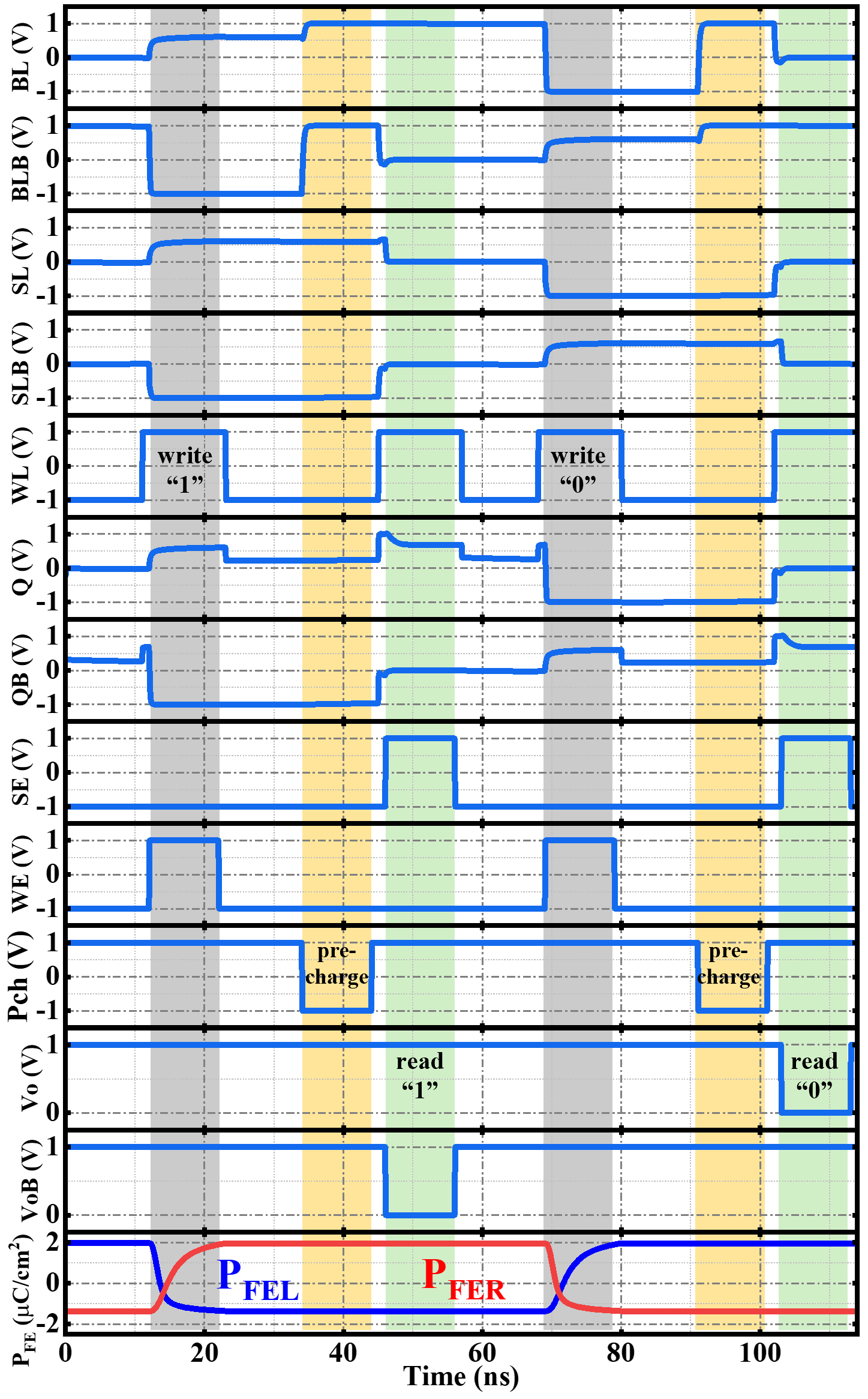}}
    \caption{Waveform of signals applied to the memory circuit in Fig.~\ref{fig:array-and-layout}. The bit-cell will first be written to ``1'' followed by read operation and then written to ``0'' followed by another read operation.}
    \label{fig:signals}
\end{figure}

Transient simulation is performed to analyze our 4T differential memory bit-cell.
The waveforms of the various signals are shown in Fig.~\ref{fig:signals}.
In this simulation, the write pulse width and magnitude are $\textit{t}_\text{p}=10$~ns and $\textit{V}_\text{w}=2$~V, respectively, to ensure the bit-cell is successfully written into.

In Fig.~\ref{fig:signals}, the bit-cell goes through two write-read cycles.
The bit-cell is first written with ``1'' before a read operation is applied to check the state of the bit-cell.
The bit-cell is then overwritten with ``0'' before a read operation is performed to check the state of the bit-cell again.

As Fig.~\ref{fig:signals} shows, during the write ``1'' operation (left most grey-shaded region), $\textit{V}_\text{BL}=0.6~\text{V}$ and $\textit{V}_\text{BLB}=-1~\text{V}$, resulting in switching of $P_\text{FER}$ and $P_\text{FEL}$.
The read operation consist of a pre-charge step and a sensing step. 
During pre-charge step (left most yellow-shaded region in Fig.~\ref{fig:signals}), Pch is grounded, \ac{WL} is disabled, and $\textit{V}_\text{BL}=\textit{V}_\text{BLB}=1~\text{V}$.
After the pre-charge step is completed, \ac{WL} is activated and the SA is enabled to sense the state of the selected bit-cell (left most green-shaded region in Fig.~\ref{fig:signals}).
$Q$ remains at a high-level voltage compared with $QB$ since $P_\text{FEL}=-P_\text{FE}$ and $P_\text{FER}=+P_\text{FE}$, resulting in different discharging time between $Q$ and $QB$.
The waveforms for Vo and VoB suggest that the state stored in the bit-cell is indeed ``1''.
Next, ``0'' is wrtten into the bit-cell (right most gray-shared region in Fig.~\ref{fig:signals}) followed by the read operation.
The waveforms for $P_\text{FEL}$ and $P_\text{FER}$ indicate the write operation is successful.
The final readout is also correct.
% During the write ``0'' operation (right most gray-shared region in Fig.~\ref{fig:signals}), $P_\text{FER}$ and $P_\text{FEL}$ are successfully switched.
% The read operation that follows validates the successfully reading out of ``0'' from the bit-cell.
Thus, the functionality of our proposed 4T differential \ac{FeFET} bit-cell is validated.
 
% As the figure shows, after the write pulses are applied on DBL, DSL, DBLB, and DSLB, the \ac{BL} and BLB are changed to a differential voltage level, leading to a differential voltage drop between $Q$ and $QB$.
% The voltage drop between $Q$ and $QB$ will assist to write the state to the polarization of the cross-coupled \ac{FeFET}s.
% The polarization of the left \ac{FeFET}, $P_\text{FEL}$ is changed to $-P_\text{FE}$ and the polarization of the right \ac{FeFET}, $P_\text{FER}$ is changed to $+P_\text{FE}$ during write ``1'' cycle.
% After the write operation is completed, BL and BLB is pre-charged to $V_\text{read}=0.2$ V.
% Then the sense amplifier is enabled to sense the differential voltage on BL and BLB.
% The red rectangular with the zoomed in area in Fig.~\ref{fig:simulation-results-transient} indicates the changes of the outputs from the sense amplifier.
% The needed time for Vo to be discharged from high to low is defined as sense delay for the sense amplifier and is used as read latency for the bit-cell evaluation.

% From the transient simulation, the voltage on BL and BLB, Q and QB during read operation, the polarization of the cross-coupled FeFETs ($P_\text{FEL}$ and $P_\text{FER}$) during read operation will be extracted for further design space exploration which will be discussed as follow.

% \subsection{Simulation Waveforms}

% Fig.~\ref{fig:signals} shows the signals waveform applied to the 4T differential memory in our simulation.

% The bit-cell go through two write-read cycles.
% As the figure shows, in the first cycle, the bit-cell is written to ``1'' by applying $V_\text{w}/2$ pulses on DBL and DSL and $-V_\text{w}/2$ pulses on DBLB and DSLB.
% This write operation is followed by a read operation: PCH is lowering to GND to pre-charge and equalize the \ac{BL} and BLB, then SE is on to enable the sense amplifier, Vo and VoB will output the corresponding voltages that read from the 4T differential memory bit-cell.
% After the first write-read cycle complete, the bit-cell go through the second cycle by writing the cell to ``0'', i.e. $-V_\text{w}/2$ pulses on DBL and DSL and $V_\text{w}/2$ pulses on DBLB and DSLB.
% The read operation is the same as the first cycle.

\subsection{Design Space Exploration}

\begin{figure}[!b]
    \centerline{\includegraphics[width=\linewidth]{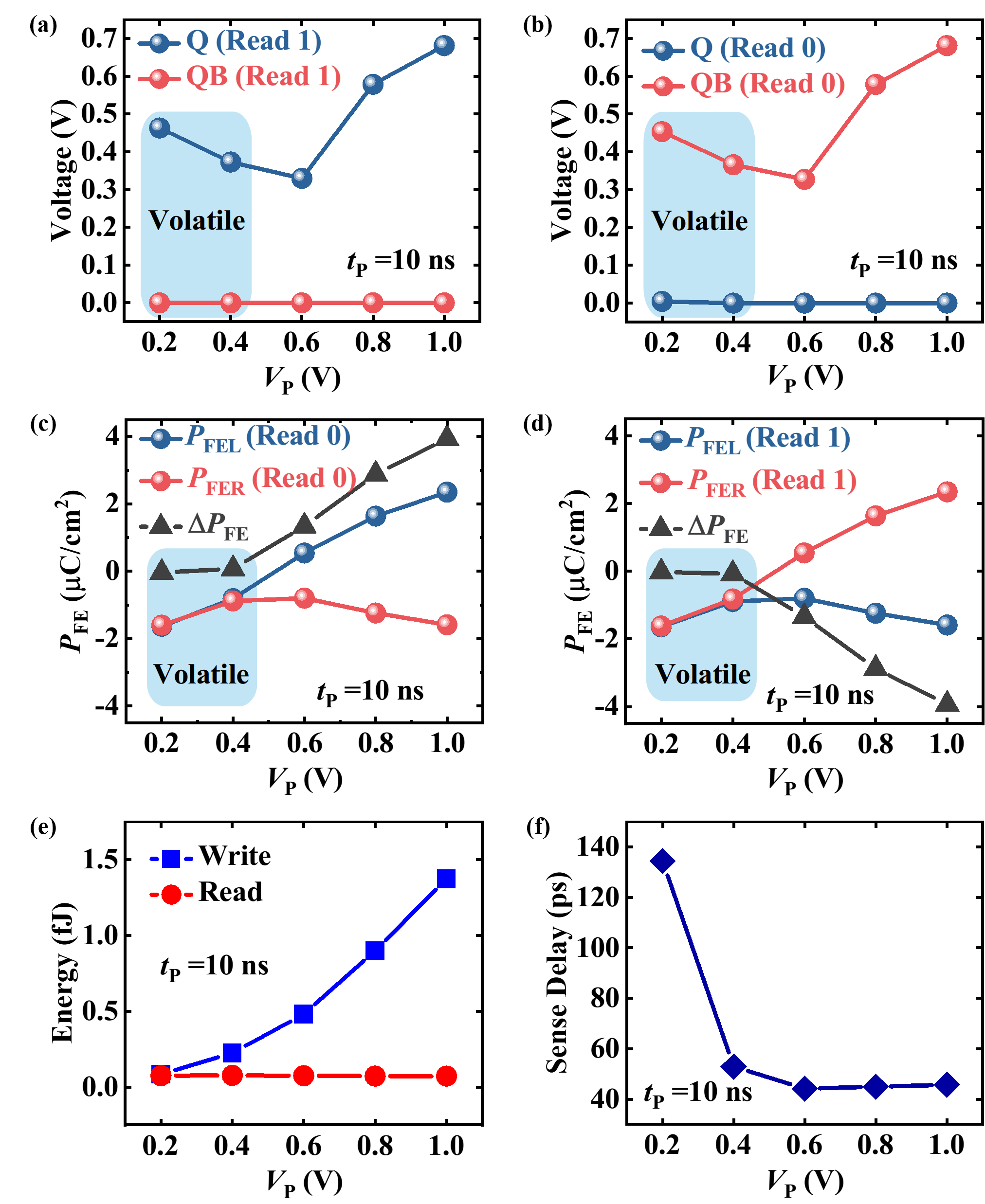}}
    \caption{Proposed 4T differential bit-cell under different write conditions with $\textit{t}_\text{p}=10~\text{ns}$ and $\textit{V}_\text{p}=\textit{V}_\text{w}/2$ varying from $0.2$~V to $1$~V in $0.2$~V steps. (a) and (b) are the voltages of $Q$ and $QB$ during read ``1'' and read ``0'', respectively. (c) and (d) are the polarization of the \ac{FE} layer of the FeFETs ($P_\text{FEL}$ and $P_\text{FER}$) and the difference of $P_\text{FEL}$ and $P_\text{FER}$ during read ``1'' and read ``0'', respectively. (e) The write and read energy under different write conditions. (f) Sensing delays under different write conditions.}
    \label{fig:bitcell-point-voltage}
\end{figure}

The write conditions can affect the performance and functionality of the proposed 4T differential \ac{FeFET} bit-cell. %, resulting in different design specifications due to different operating modes of the bit-cell.
To further evaluate the performance of the proposed 4T differential \ac{FeFET} bit-cell, we vary the control signals that are applied to the bit-cell.
We focus on different write conditions in this section.
The read condition is fixed: the pre-charge voltage, $\textit{V}_\text{read}$, and the sensing time are kept at $1$~V and $10$~ns, respectively.
%The write voltage $V_\text{w}$ is varied from 0.4 V to 2 V and the write pulse width $t_\text{p}$ is varied from 2 ns to 10 ns.
Fig.~\ref{fig:bitcell-point-voltage} shows the bit-cell performance with $\textit{t}_\text{p}=10$~ns.
$\textit{V}_\text{p}=\textit{V}_\text{w}/2$ is varied from $0.2$~V to $1$~V in $0.2$~V steps to study the bit-cell behavior.

% \begin{figure*}[!h]
%     \centerline{\includegraphics[width=\linewidth]{images/bitcell_10ns.png}}
%     \caption{Proposed 4T differential bit-cell under different write conditions with $t_\text{p}$=10 ns and $V_\text{p}=V_\text{w}/2$ varying from 0.2 V to 1 V with step of 0.2 V. (a) and (b) are the voltage on BL and BLB during read ``1'' and read ``0'' respectively, (c) and (d) are the voltage on $Q$ and $QB$ during read ``1'' and read ``0'' respectively, (e) and (f) are the polarization of the \ac{FE} layer of the FeFETs ($P_\text{FEL}$ and $P_\text{FER}$) and the difference of $P_\text{FEL}$ and $P_\text{FER}$ during read ``1'' and read ``0'' respectively, (g) write and read energy under different write conditions, (h) sense delay under different write conditions.}
%     \label{fig:bitcell-point-voltage}
% \end{figure*}

Fig.~\ref{fig:bitcell-point-voltage} (a) and (b) show $\textit{V}_{Q}$ and $\textit{V}_{QB}$ during read ``1'' and read ``0'', respectively, after different write conditions.
It can be clearly observed that $\textit{V}_{Q}$ and $\textit{V}_{QB}$ decreases when  $\textit{V}_\text{p}\leq0.6~\text{V}$ and increases when $\textit{V}_\text{p}\geq0.6~\text{V}$ during read ``1'' and during read ``0'', respectively.
To explain this trend under different write conditions, the polarizations of the \acp{FeFET}, $P_\text{FEL}$ and $P_\text{FER}$, are extracted as shown in Fig.~\ref{fig:bitcell-point-voltage} (c) and (d).
For $\textit{V}_\text{p}=0.2~\text{V}$ and $\textit{V}_\text{p}=0.4~\text{V}$, the difference between $P_\text{FEL}$ and $P_\text{FER}$ ($\Delta P=P_\text{FEL} - P_\text{FER}$) is around 0 \si{\mu C/cm^2}, because $P_\text{FEL}=P_\text{FER}=-P_\text{FE}$ under both read ``0'' and read ``1'' conditions as shown in Fig.~\ref{fig:bitcell-point-voltage} (c) and (d), which indicates a write operation failure.
However, $Q$ and $QB$ can still hold the difference during the read operation because when both FeFETs are at $-P_\text{FE}$, they can be considered as a high $\textit{V}_\text{th}$ NMOS and the data is stored as charge on the parasitic capacitances at $Q$ and $QB$.
When the read operation is applied, the $\textit{V}_\text{gs}$ of NF1 is smaller than that of NF2, leading to different discharge speed of BL and BLB---NF1 discharges $Q$ slower NF2 discharges $QB$ and therefore, the difference between BL and BLB can be obtained.
Therefore, when $\textit{V}_\text{p}\leq0.4~\text{V}$, the proposed 4T differential \ac{FeFET} bit-cell is operating as volatile memory. 
When $\textit{V}_\text{p}\geq0.6~\text{V}$, $|\Delta P|$ is larger than 0 \si{\mu C/cm^2} and the \ac{FE} layers in the bit-cell were successfully programmed.
Thus, the bit-cell operates as non-volatile memory.
Hence, the bit-cell can operate as either volatile or non-volatile memory by adjusting the write condition of the bit-cell.
Fig.~\ref{fig:bitcell-point-voltage} (e) shows the write and read energy with different $\textit{V}_\text{p}$.
The write energy increases as $\textit{V}_\text{p}$ increases whereas the read energy is not strongly affected by the write conditions.
Fig.~\ref{fig:bitcell-point-voltage} (f) shows the sensing delay under different write conditions.
The sensing delay reaches to 137~ps at $\textit{V}_\text{p}=0.2~\text{V}$ whereas the sensing delay is around 47~ps under other conditions.
As shown in Fig.~\ref{fig:bitcell-point-voltage}, the proposed differential bit-cell can be successfully programmed as long as the write voltage is larger than 0.4 V, indicating the bit-cell does not require precise control for the write operation.

\begin{figure}[!t]
    \centerline{\includegraphics[width=\linewidth]{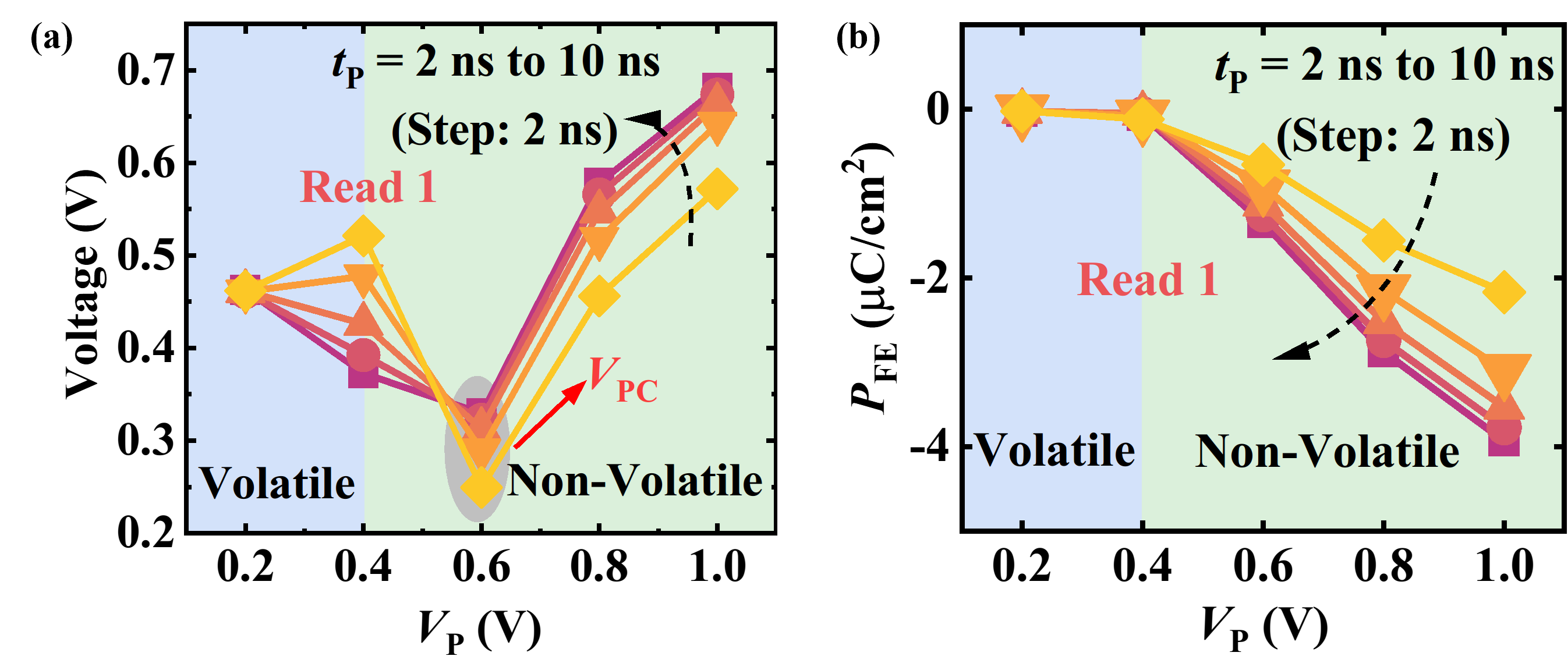}}
    \caption{Memory window at different write conditions. (a) difference between $Q$ and $QB$ ($\Delta Q=\textit{V}_{Q}-\textit{V}_{QB}$) during read ``1'' at different write conditions with different $V_\text{p}$ and $t_\text{p}$. (b) difference between $P_\text{FEL}$ and $P_\text{FER}$ ($\Delta P=P_\text{FEL} - P_\text{FER}$) during read ``1'' at different write conditions.}
    \label{fig:delta-different-write-condition}
\end{figure}

%From Fig.~\ref{fig:bitcell-point-voltage} (a) to (d), $V_\text{p} = \text{0.6 V}$ is a critical voltage for $t_\text{p} = \text{10 ns}$ which separate the volatile mode and the non-volatile mode.
To further explore the design space of the differential \ac{FeFET} bit-cell, we include the $\textit{t}_\text{p}$ variation at the different $\textit{V}_\text{p}$.
Fig.~\ref{fig:delta-different-write-condition} shows the \acp{MW} under different write conditions.
We define the extrinsic \ac{MW} and intrinsic \ac{MW} as the difference between $\textit{V}_{Q}$ and $\textit{V}_{QB}$ (\textit{i.e.}, $\Delta Q=\textit{V}_{Q}-\textit{V}_{QB}$) and $\Delta P=P_\text{FEL} - P_\text{FER}$, respectively.
Fig.~\ref{fig:delta-different-write-condition}(a) shows $\Delta Q$ under different write conditions during read ``1''.
The trend is the same as Fig.~\ref{fig:bitcell-point-voltage} (a)-(b), where $\Delta Q$ decreases when $\textit{V}_{p} \leq 0.6~\text{V}$ and increase when $\textit{V}_{p} \geq 0.6~\text{V}.$
When $\textit{V}_\text{p}=0.4~\text{V}$ and $\textit{t}_\text{p}\leq4~\text{ns}$, $\Delta Q$ is higher than when $\textit{V}_\text{p}=0.2~\text{V}$ due to the charge pumping effect when parasitic capacitance is small (the bit-cell parasitic capacitance is 58~aF).
When $\textit{V}_\text{p}\geq0.6~\text{V}$, $\Delta Q$ increases with both $\textit{V}_\text{p}$ and $\textit{t}_\text{p}$.
The same trend can be seen in Fig.~\ref{fig:delta-different-write-condition}(b) as in Fig.~\ref{fig:bitcell-point-voltage}(d), that the absolute value of $\Delta P$ increases with both $\textit{V}_\text{p}$ and $\textit{t}_\text{p}$.
When $\textit{V}_\text{p}=0.4~\text{V}$, $\Delta Q$ increases as $t_\text{p}$ increases.
This is because $\Delta P$ is smaller at $\textit{t}_\text{p}=2~\text{ns}$ as compared to other conditions according to Fig.~\ref{fig:delta-different-write-condition}(b).
$\Delta Q$ for different $\textit{t}_\text{p}$ under $\textit{V}_\text{p}=0.2~\text{V}$ are the same, due to same $\Delta P$.
Therefore, $\textit{V}_\text{p}=0.6~\text{V}$ is the critical voltage to configure the cell between volatile and non-volatile memory.
The results shown in Fig.~\ref{fig:bitcell-point-voltage} and Fig.~\ref{fig:delta-different-write-condition} indicate that the proposed bit-cell maintains stable operation across a wide range of write voltages and pulse widths.
% \begin{figure}[!h]
%     \centerline{\includegraphics[width=0.5\linewidth]{images/bitcell-vary-tp-write-energy.png}}
%     \caption{Write energy at different write conditions with different $V_\text{p}$ and $t_\text{p}$.}
%     \label{fig:energy-different-write-condition}
% \end{figure}

% Fig.~\ref{fig:energy-different-write-condition} shows the write energy under different write conditions.
% The write energy is proportional to the write pulse width and write voltage.

% \begin{figure}[!h]
%     \centerline{\includegraphics[width=\linewidth]{images/bitcell-vary-tp-read.png}}
%     \caption{(a) Read energy under different write conditions, (b) sense delay under different write conditions.}
%     \label{fig:read-different-write-condition}
% \end{figure}

% Fig.~\ref{fig:read-different-write-condition} shows the read performance under different write conditions with different $V_\text{p}$ and $t_\text{p}$.
% As the Fig.~\ref{fig:read-different-write-condition}(a) shows, the read energy does not show strong relation with different $V_\text{p}$ and $t_\text{p}$.
% The read energy is around 0.075 fJ under all write conditions in this work.
% The sense delay shows the same trend as Fig.~\ref{fig:bitcell-point-voltage}(h).
% The sense delay reach maximum to about 137 ps at $V_\text{p}=\text{0.2}$ V and not affected by the write pulse width.

% \begin{figure}[!h]
%     \centerline{\includegraphics[width=\linewidth]{images/performance.png}}
%     \caption{Caption}
%     \label{fig:enter-label}
% \end{figure}

\begin{table}[!h]
\caption{Comparison of our proposed 4T differential \ac{FeFET} bit-cell with previous designs and gain cell deisgn}
\resizebox{\linewidth}{!}{\begin{tabular}{|l|l|l|l|l|l|l|}
\hline
Design         & 6T-SRAM\cite{sle10t}  & 4T-R\cite{nvsram4t}   & 7T2R\cite{7t2r} & 8T2R\cite{8t2r}  & 1T1FeFET\cite{1t1fe} & \textcolor{red}{This work}    \\ \hline
\# Transistors & 6        & 4            & 7          & 8 & 2           & \textcolor{red}{4}            \\ \hline
Type & Simulation & Simulation & Experiment & Simulation & Experiment & \textcolor{red}{Simulation} \\ \hline
NVM device     & N.A.     & R-FeFET      & ReRAM      & MTJ & FeFET & \textcolor{red}{FeFET}        \\ \hline
Store voltage  & 0.74 V   & 1.1 V        & 1.8 V      & 1.2 V  & 0.9 V     & \textcolor{red}{2 V}          \\ \hline
Store power    & 37.12 \si{\mu W} & 4.45 mW      & 0.311 mW   & 0.604 \si{\mu W}  & 0.134 \si{\mu W}  & \textcolor{red}{0.13 \si{\mu W}}      \\ \hline
Store time     & 1.92 ns  & N.A.        & 10 ns      & 2 ns     & 20 ns   & \textcolor{red}{2 ns}          \\ \hline
Sense Scheme & Differential & Differential & Differential & Differential & Current-Based & \textcolor{red}{Differential} \\ \hline
B\&E           & N.A.     & Not required & Required   & Required   & N.A.   & \textcolor{red}{Not required} \\ \hline

\end{tabular}}
\label{tab:benchmark}
\end{table}

Our proposed 4T differential memory bit-cell is compared with previous designs in Table~\ref{tab:benchmark}.
Compared to conventional SRAM and other nvSRAM designs, our design requires only four transistors.
The FeFET structure is CMOS-compatible, which eliminates the need for specialized fabrication processes as described in \cite{nvsram4t}.
Additionally, the store power is 0.13~\si{\mu W}, which is exceptionally low compared to \cite{8t2r}.
The lower store power is mainly attributed to the write scheme of the bit-cell. During writing, BL and SL are driven to the same voltage, and BLB and SLB are driven to the same voltage, thereby reducing the potential difference across these nodes and minimizing the leakage current.
Since the proposed differential bit-cell is also a cross-coupled gain cell, we compared it with a state-of-the-art 1T–1FeFET gain cell \cite{1t1fe}. The proposed design achieves comparable store power, while adopting a differential read scheme that can be directly integrated with standard SRAM sensing circuits.
% The store time is 2~ns, which is comparable to 6T SRAM.
Owing to the differential topology, the proposed bit-cell achieves a 2~ns store time, comparable to 6T SRAM, since correct sensing only requires a sufficient polarization-state difference between the two FeFETs.
For reliability and endurance concerns, previous studies have demonstrated excellent FeFET reliability and endurance (e.g., retention over $10^4$ s at 85\si{\celsius} and endurance up to $10^{12}$ cycles reported in \cite{1t1fe}). The proposed bit-cell can directly benefit from such device-level improvements without requiring any modification to its circuit structure.

%% file: conclusion.tex
\section{Conclusion}
% The conclusion goes here.
In this work, we propose a FeFET-based differential non-volatile memory bit-cell. The proposed bit-cell comprises a pair of 2T gain cells connected in a cross-coupled topology, and uses only four transistors in total, thereby reducing the cell area compared with conventional CMOS-based SRAM and prior nvSRAM designs. By adjusting the write conditions, the bit-cell can be configured to operate in either volatile or non-volatile mode. In the non-volatile mode, the store power is 0.13~\si{\mu}W with a 2~ns store time, which is reduced compared with previous designs. Moreover, the proposed design does not require explicit backup-and-restore (B\&R) operations.